\documentclass[sigconf,natbib=true,nonacm]{acmart}

\usepackage{algorithm}
\usepackage{algorithmic}

\usepackage{multirow}

\AtBeginDocument{%
  \providecommand\BibTeX{{%
    \normalfont B\kern-0.5em{\scshape i\kern-0.25em b}\kern-0.8em\TeX}}}

\setcopyright{acmcopyright}
\copyrightyear{2018}
\acmYear{2023}
\acmDOI{XXXXXXX.XXXXXXX}

\acmConference[SIGIR 2023]{The 46th International ACM SIGIR Conference on Research and Development in Information Retrieval}{July 23-27, 2023}{Taipei, Taiwan}
\acmPrice{15.00}
\acmISBN{978-1-4503-XXXX-X/18/06}

\begin{document}

\title{Light-weight End-to-End Graph Interest Network for CTR Prediction in E-commerce Search}




\author{Pipi Peng}
\email{ppskingdom@gmail.com}
\affiliation{%
  \city{Shanghai}
  \country{China}
}

\author{Yunqing Jia}
\authornote{This author is the one who gives a lot of guidance in the work.}
\email{jiayunqing01@hotmail.com}
\affiliation{%
  \city{Shanghai}
  \country{China}
}

\author{Ziqiang Zhou}
\email{ziqiang0711@gmail.com}
\affiliation{%
  \city{Shanghai}
  \country{China}
}

\author{murmurhash}
\email{murmurhash@outlook.com}
\affiliation{%
  \city{Shanghai}
  \country{China}
}

\author{Zichong Xiao}
\email{zichong.xiao@gmail.com}
\affiliation{%
  \city{Shanghai}
  \country{China}
}


\begin{abstract}
Click-through-rate (CTR) prediction has an essential impact on improving user experience and revenue in e-commerce search. The ambiguous and incomplete nature of search queries makes it important for CTR models to mine users' search intentions from rich historical behavior data. With the development of deep learning, a series of works has been proposed to model user interests, bringing significant improvement in model performance. Attention-based methods focus on summarizing user behaviors into a comprehensive interest representation depending on their relationship with the target. Graph-based methods are well exploited to utilize graph structure extracted from user behaviors and other information to help embedding learning. However, most of the previous graph-based methods face the challenges of deployment and performance in large-scale e-commerce search systems. First, these methods usually require a separate graph engine for graph storage and sampling, which makes it hard to jointly train graph embedding with CTR prediction, while requiring more implementation effort. Second, they mainly focus on recommendation scenarios, and therefore their graph structures highly depend on item's sequential information from user behaviors, ignoring query's sequential signal and query-item correlation. In both practice and our experiments, this extra information brings notable improvement because of the query-dependent nature of e-commerce search.

In this paper, we propose a new approach named Light-weight End-to-End Graph Interest Network (EGIN) to effectively mine users' search interests and tackle previous challenges. (i) EGIN utilizes query and item's correlation and sequential information from the search system to build a heterogeneous graph for better CTR prediction in e-commerce search. (ii) EGIN's graph embedding learning shares the same training input and is jointly trained with CTR prediction, making the end-to-end framework effortless to deploy in large-scale search systems. The proposed EGIN is composed of three parts: query-item heterogeneous graph, light-weight graph sampling, and multi-interest network. The query-item heterogeneous graph captures correlation and sequential information of query and item efficiently by the proposed light-weight graph sampling. The multi-interest network is well designed to utilize graph embedding to capture various similarity relationships between query and item to enhance the final CTR prediction. We conduct extensive experiments on both public and industrial datasets to demonstrate the effectiveness of the proposed EGIN. At the same time, the training cost of graph learning is relatively low compared with the main CTR prediction task, ensuring efficiency in practical applications. Our code will be publicly available.
\end{abstract}


\keywords{sponsored search, click-through rate prediction, graph neural network, behavior sequence}

\maketitle
\pagestyle{plain}

\section{Introduction}

In sponsored search, click-through rate (CTR) prediction is vital for improving user experience and total revenue. The accuracy of CTR prediction depends on the system's ability to understand users'  current search intentions and historical interests. The ambiguity of search queries and the consistency of user interests encourage search systems to exploit user behavior in a more and more elaborate way.

With the development of search and recommender systems, user behavior sequence is inspected in different ways to mine user interest. Recently, a series of works focusing on modeling latent user interest from historical behaviors use various deep neural network architectures including CNN \cite {tang_topn, yuan_gen}, RNN \cite{dsin, zhou2018deep}, Transformer \cite {chen2019behavior, masked_transformer}, Capsule \cite {li_mind}, etc. Besides, Graph Neural Network \cite{perozzi2014deepwalk, GCN, wang2018billion, li2019graph, KGAT} has become a popular method for embedding learning to aid CTR prediction in search and recommendation systems and achieved great success. The learned graph embedding is usually consumed by the CTR prediction network to combine the graph learning and the final CTR objective.

These graph-based methods, although following either a two-stage or end-to-end approach, still require first constructing a graph and then sampling using a graph engine. We summarize them as \textbf{Construct\&Sample} paradigm. When applying these methods in a large-scale e-commerce system with billions of items and samples, updating and retrieving the graph within a short delay can be challenging, thus becoming the bottleneck of end-to-end joint training of graph embedding with CTR prediction task.

In addition, these graph-based methods mentioned above mainly focus on items' sequential and attributive information during graph construction, ignoring queries' sequential information and query-item correlation, which help improve search ctr prediction quality in our industrial practice.

We propose a new approach called \textbf{Light-weight End-to-End Graph Interest Network (EGIN)} to solve these problems. Firstly, the resource consumption and time delay of the Construct\&Sample paradigm is alleviated by introducing a light-weight graph sampling method based on data manipulation of CTR task input instead of building a graph engine. Secondly, the EGIN utilizes the query-item heterogeneous graph for search CTR prediction.

The main contributions of this paper are as follows.
\begin{itemize}
\item We introduce our light-weight graph sampling that shares the same training input with the CTR prediction task without physically storing the graph structure. End-to-end joint training of graph embedding and CTR prediction is implemented without reorganizing complicated graph data or depending on the graph engine. Our method can be effortlessly integrated into other CTR prediction networks.
\item We propose the query-item heterogeneous graph to model query-item correlation in graph structure, which improves CTR prediction performance in search scenarios compared with the item-only version. We also design the multi-interest network to exploit query-item correlation provided by graph learning toward a better understanding of user interest.  
\item We conduct extensive experiments on public and industrial datasets to demonstrate the effectiveness of the proposed EGIN framework. The online A/B test is conducted to verify the productive performance of the proposed approach. We also discuss the influence of different training techniques and data management. 
\end{itemize}

\section{Related Work}

\subsection{CTR Prediction}
CTR prediction has gained attention from researchers for many years because of its vital role in search and recommendation systems and its ability to improve the revenue of online applications largely. Considering the high sparsity of the input features, a series of works have been proposed to capture feature interactions. Factorization Machines (FM) \cite{FactorizationMachines} uses a low-dimensional vector for feature representation and learns the second-order crossover of features by the inner product. Wide\&Deep \cite{widendeep} jointly trains the wide linear unit for and the deep MLP layer to enhance both memorization and generalization. DeepFM \cite{deepFM} integrated factorization machines and deep neural networks to learn the low-order feature interactions. xDeepFM \cite{xDeepFM} proposes a novel Compressed Interaction Network (CIN) to model high-order feature interactions in an explicit fashion and adopts the traditional DNN simultaneously. Deep \& Cross network (DCN) \cite{DCN} adopts interaction with representation in each layer with original feature embedding to learn higher-order feature representations. Deep learning-based methods have also achieved great success in user interest mining to aid CTR prediction. Deep Interest Network (DIN) \cite{zhou2018deep} developed an attention-based method to assign different weights to historical commodities according to their relationship with the target commodity. Depp Interest Evolution Network (DIEN \cite{zhou2019deep} further utilizes RNN to model the evolution of user interest by taking sequential information into account. Deep Session Interest Network (DSIN) \cite{dsin} leverages Bi-LSTM with self-attention layers to capture users' inter-session and intro-session interests. MIMN \cite{pi2019mimn} tackles the challenge of long sequential user behavior modeling by decoupling the user interest model from the entire framework and designing the User Interest Center (UIC) to record new behaviors incrementally. Behavior Sequence Transformer (BST) \cite{chen2019behavior} uses Transformer to capture underlying sequential signals from user behavior for a better recommendation.

\subsection{Graph Neural Networks for Search and Recommendation}
A group of graph embedding methods is introduced into the CTR prediction task with their strong potential for modeling graph information in search and recommendation. EGES \cite{wang2018billion} adopts DeepWalk \cite{perozzi2014deepwalk} to construct graphs based on click sequence, and the Skip-Gram model is used for graph embedding learning. Then the learned node representation is consumed in the CTR prediction network. Graph Intention Network (GIN) \cite{li2019graph} utilizes user behavior sequence to build a co-occurrence commodity network and applies graph diffusion and aggregation to enrich node representation of historical clicks to overcome behavior sparsity and weak generalization problems. KGAT \cite{KGAT} combines user-item graph with knowledge graph for collaborative knowledge and then applies graph convolution to get the graph representation. Heterogeneous graph Attention Network (HGAT) \cite{HGAT} utilizes a semantic-level and a node-level attention network to discriminate the importance of neighbor nodes and node types.  DG-ENN \cite{DG-ENN} uses attribute graph and collaborative graph to refine the embedding with strategies, and alleviates the feature and behavior sparsity problems. However, when applying this method directly in CTR prediction in an end-to-end manner, a separate graph engine is needed to restore and sample the graph for embedding learning. Otherwise, the two-stage approach which first conducts graph embedding learning and then consumes in CTR prediction brings the problem that the graph representation is not optimized for the final objective. Also, these methods focus more on recommendations and ignore the potential of capturing interactions between items and queries to build heterogeneous graphs. To face these challenges, we propose a novel approach to conduct high-efficient end-to-end graph learning in e-commerce search scenarios.

\section{The Proposed Approach}

\begin{figure*}[h]
  \centering
  \includegraphics[width=0.86\linewidth]{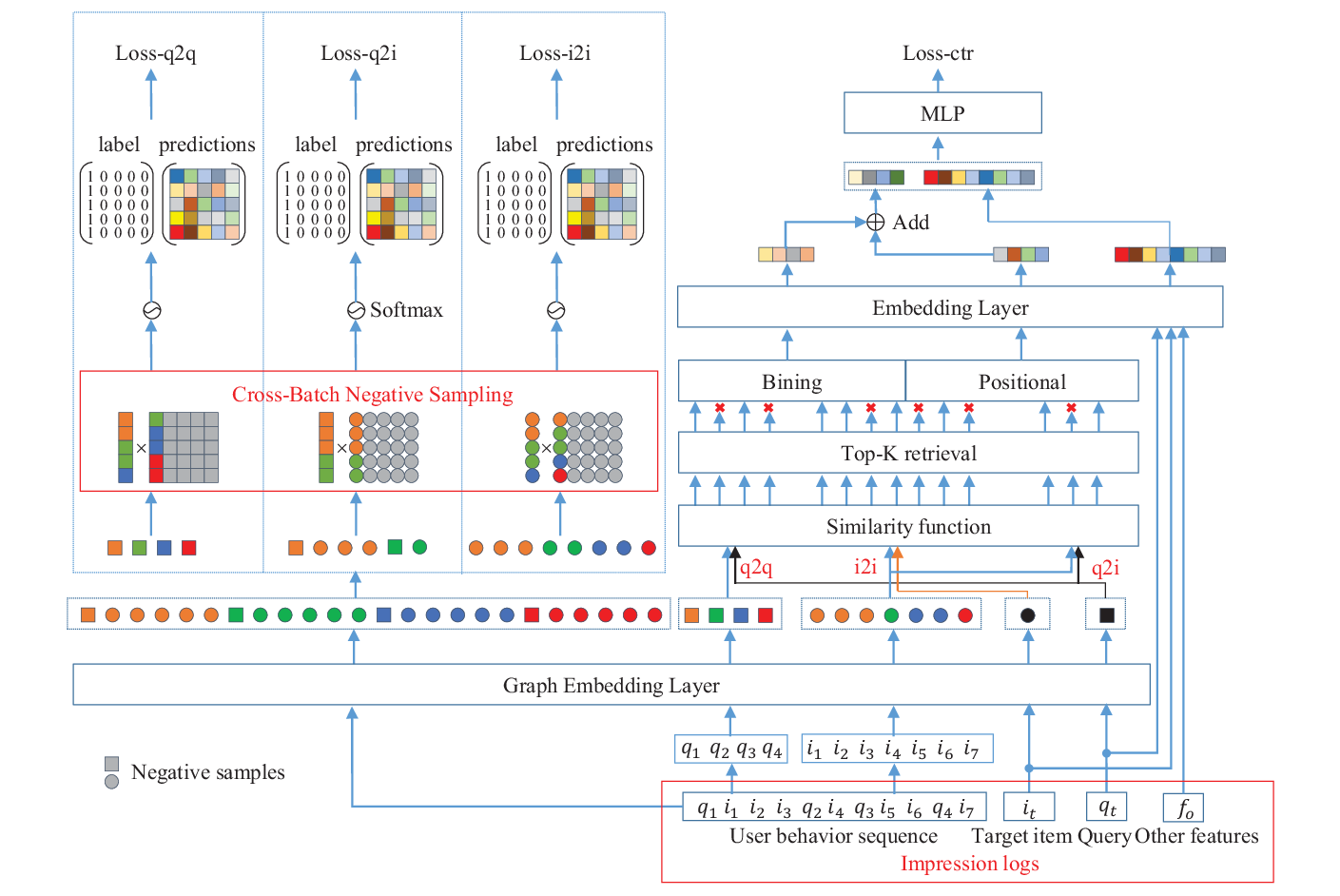}
  \caption{The framework of the proposed EGIN model. We use the unified input for both graph learning and CTR prediction network. Our query-item heterogeneous graph is constructed based on user behavior sequence and conducts embedding learning based on edges. We build i2i edges for every neighbor within the distance of 2 in click sequence and build q2i/q2q pairs by time and category constraints. The same behavior sequence is provided to the CTR prediction network, where multiple similarity relationships are calculated based on graph embedding.}
  \Description{The framework of EGIN. Left-side shows how the graph is structured as pairs, right-side shows how features are fed into the CTR prediction model.}
  \label{fig:framework}
\end{figure*}

In this section, we introduce our Light-weight End-to-End Graph Interest Network (EGIN) in detail. As shown in Figure~\ref{fig:framework}, EGIN is composed of two parts, on the left side is the query-item heterogeneous graph, and on the right side is the multi-interest network. Notice that the two parts are jointly optimized in an end-to-end manner and share the same input extracted from search impression logs. Equation~\ref{eq:loss} presents our joint training objective.

\subsection{Query-item Heterogeneous Graph}

\begin{figure}[h]
  \centering
  \includegraphics[width=\linewidth]{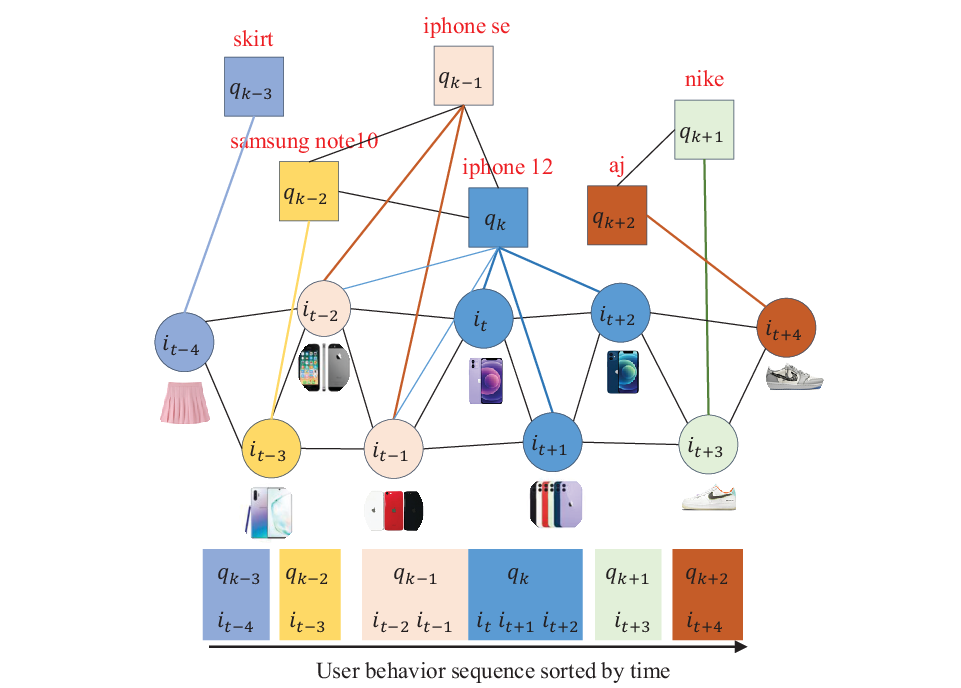}
  \caption{Query-item heterogeneous graph structure. User behavior sequence containing query and click sequence ordered in time. First, neighboring items are connected by a window size of 2. Then queries are segmented into sessions based on their semantic similarity and happening time. Queries within the same session are all connected. Finally, every query and its nearby items within the time window that satisfy categorical constraints with the query is connected to capture item-query correlation.}
  \Description{Graph structure.}
  \label{fig:graph}
\end{figure}

In previous graph-based CTR prediction work, users' historical behaviors are well explored to represent personal interest. In e-commerce search scenarios, queries explicitly describe interest evolution and current search intention. However, previous work mainly focuses on semantic knowledge conducted by the query, ignoring its sequential information and correlation with items. 

To jointly model signals of click sequence and query sequence extracted from user behavior, we introduce our query-item heterogeneous graph as shown in Figure~\ref{fig:graph}. With query and item as nodes, our graph architecture contains three kinds of edges: \textbf{item2item}, \textbf{query2query} and \textbf{query2item}:

\textbf{item2item}. Following previous work, we utilize user click sequence to capture item2item relationship based on co-occurrence following Skip-Gram \cite{mikolov2013efficient, mikolov2013distributed}. For example, in Figure~\ref{fig:graph}, items are connected according to the window size. Notice that instead of splitting the click sequence into sessions by time frame or user interest, we adopt the original click sequence to learn the item2item relationship. 

User's click sequence usually forms a list of several clusters, where items within the cluster are often similar to each other. In contrast, cross-cluster items are quite different in the category, representing the user's interest transfer. Considering these two properties, user's click sequence can provide information about both intro-cluster similarity and cross-cluster transfer probability. 

\textbf{query2query}. We adopt users' historical query sequences to build the query2query relationship. Query sequence is segmented into multiple sessions similar to \cite{dsin} based on semantic similarity and time constraint to prevent edge construction across dissimilar queries. Queries within the same session are treated as similar and are connected with each other. As in Figure~\ref{fig:graph}, all queries form three sessions, and each session constructs a complete graph.

\textbf{query2item}. After capturing sequential information of item and query separately, the relationship between query and item is preserved in our graph to jointly learn their embeddings, which plays an important role in modern search systems. For each query in a user's behavior sequence, the clicked items within a time window before or after this query are filtered by category constraints to become the relevant items. Then we connect edges between every query and its relevant items. As in Figure~\ref{fig:graph},  queries and their relevant items are painted the same colors, but only relevant items are connected with corresponding queries.

\subsection{Light-weight Subgraph Sampling}
When adopting graph-based methods in CTR prediction, previous work usually constructs graphs based on global or historical information, and the whole graph is stored physically. During embedding learning, the subgraph is sampled from the whole graph to fit in the memory. We summarize them as \textbf{Construct\&Sample} paradigm. When applying these methods in industrial scenarios with billions of items and samples, updating and retrieving graphs with a short delay can be challenging, thus becoming the bottleneck of end-to-end joint training of graph embedding and CTR prediction task.

Towards a light-weight subgraph sampling method, we first consider an underlying graph representing current global user interest as \textbf{implicit graph}. Statistically, every user's mini-graph is merged together to form the implicit graph. Previous graph construction methods based on accumulative data are gradual approximations of the implicit graph. In this article, we consider user behavior sequence or its slightly manipulated version to be a subgraph sampling of the implicit graph, thus avoiding storing and sampling the graph physically.

The pseudo-code of the light-weight graph sampling and learning method is listed in Algorithm~\ref{alg:embeddinglearning}. Given users' click sequences and query sequences from impression log data, we first build edges between items with co-occurrence according to the window size. Then query2query and query2item edges are connected using the BuildGraph method described in Algorithm~\ref{alg:BuildEdge}, explaining how two entities are related according to time and category constraints. Finally, our training objective over graph embedding is computed based on all the edges. With billions of items and queries in our dataset, we adopt Negative Sampling and softmax loss to approximately optimize the graph embedding. 
$$L_G = -\mathrm{log}\frac{\mathrm{exp}(e_a\cdot e_p)}{\mathrm{exp}(e_a\cdot e_p) + \sum_{e_n \in N} \mathrm{exp}(e_a\cdot e_n)}$$ 

\begin{algorithm}
\caption{Pseudo-code of graph embedding learning}
\label{alg:embeddinglearning}
\begin{algorithmic}[1]
\REQUIRE User click sequence $C$\\
User query sequence $Q$\\
Window size $w$
\ENSURE Loss over the graph embedding
\STATE Initialization: Empty edge set $E$
\FOR {$c_j \in C$}
    \FOR {$c_k \in C[j-w:j+w]$}
        \STATE $E = E \cup \{(c_j, c_k), (c_k, c_j)\}$
    \ENDFOR
\ENDFOR
\STATE $E = E \cup \mathrm{BuildEdge}(Q, Q) \cup \mathrm{BuildEdge}(Q, C)$
\FOR{$edge$ in $E$}
    \STATE $(p_1, p_2)$ = $edge$
    \STATE $N = \mathrm{NegSample}(\mathrm{typeof}(p_2))$
    \STATE $L_{G} = \mathrm{SoftmaxLoss}(p_1,p_2,N)$
\ENDFOR
\end{algorithmic}
\end{algorithm}

\begin{algorithm}
\caption{Pseudo-code of BuildEdge}
\label{alg:BuildEdge}
\begin{algorithmic}[1]
\REQUIRE User click/query sequence $S_1$, $S_2$\\
Time mapping $\mathrm{TIME}()$\\
Category mapping $\mathrm{CAT}()$\\
timespan $T$
\ENSURE Built Edges $E$
\STATE Initialization: Empty edge set $E$
\FOR {$i_1 \in S_1$}
    \FOR {$i_2 \in S_2$}
        \IF { $ 0 < \mathrm{TIME}(i_2) - \mathrm{TIME}(i_1) < T$}
            \IF {$ \mathrm{CAT}(i_1) \cap \mathrm{CAT}(i_1) \neq \phi$}
                \STATE $E = E \cup \{(i_1, i_2), (i_2, i_1)\}$
            \ENDIF
        \ENDIF
    \ENDFOR
\ENDFOR
\end{algorithmic}
\end{algorithm}

Here $e_a$ means anchor, $e_p$ means positive sample, and $N$ represents negative samples. Graph learning loss $L_G$ can be split into $L_{i2i}$, $L_{q2q}$, and $L_{q2i}$ depending on the kind of edge from which the positive pair is extracted.

With the growth of the negative sample size, model performance also increase. In practice, we choose the negative sample number of 100 as a compromise of performance and efficiency. To support sampling for both query and item on our stream machine learning platform, we dynamically maintain a query queue and an item queue of size 1 million to conduct cross-batch negative sampling.

\subsection{Multi-interest Network for CTR Prediction}
Powered by the query-item heterogeneous graph, which jointly learns query and item embedding, we design a CTR prediction network named multi-interest network that utilizes various kinds of user interest measurements based on i2i, q2q, and q2i relationships. The network is composed of the item interest unit, query interest unit, query-item compatibility unit, and CTR prediction layer.

\textbf{Item interest unit.}
Based on graph embedding, we first compute the cosine similarity between target item $i_t$ and historical items $I = i_j (j = 1, 2, \cdots, n)$:
$$sim(i_t,i_j) = \frac{e_{i_t}^Te_{i_j}}{||e_{i_t}||\cdot||e_{i_j}||}$$
Then, we use top-k retrieval to get k historical items with the largest similarity score: $$i'_1, i'_2, \cdots, i'_k = topk(sim(i_t,I))$$ 
We adopt the equal width interval binning method \cite{dougherty1995supervised} to transformer $sim(i_t,i'_k)$ into the categorical feature and get the embedding. To capture the order information of top-k items, we add positional embedding depending on the original index of top-k items to binning embedding to get item representations:
$$e_{i'_k} = binning(sim(i_t,i'_k))  + positional(i'_k)$$
Finally, we concatenate k item representations to form the i2i feature, which is then fed into the CTR prediction network.
$$f_{i2i} = \mathrm{concate}(e_{i'_1}, e_{i'_2}, \cdots,e_{i'_k})$$
We summarize the process described in this unit as the function $\mathrm{SimExtract}(target, sequence)$, which gives a comprehensive representation of similarity between $target$ and $sequence$. Therefore i2i feature can also be written as:
$$f_{i2i} =  \mathrm{SimExtract}(i_t, I)$$

\textbf{Query interest unit.}
In addition to the item2item relationship that demonstrates whether the target item is similar to the historical items, we want to apply the same logic to query sequence to describe the relationship between queries which reveal users' explicit search interests. 

Following item interest layer, given current query $q_t$ and query sequence $Q = q_k(k=1,2,\cdots,m)$, we get q2q feature by top-k retrieval, binning and positional embedding:
$$f_{q2q} =  \mathrm{SimExtract}(q_t, Q)$$

\textbf{Query-item compatibility unit.}
Relationships between current query and historical items supplement the pure item-based or query-based interest representation and further utilize the correlation between query and items corresponding to the heterogeneous graph structure. We call this relationship query-item compatibility because it describes how well the current query is compatible with the user's interested items.

Here we also adopt the same methods to produce the compatibility information given the current query $q_t$ and historical click sequence $I$:
$$f_{q2i} =  \mathrm{SimExtract}(q_t, I)$$

\textbf{CTR prediction layer.}
We concatenate the aforementioned features $f_{i2i}$,$f_{q2q}$, $f_{q2i}$ and other features $f_o$, then feed them into an MLP layer for final CTR prediction:
$$\mathrm{pctr} = \mathrm{sigmoid}(\mathrm{MLP}(\mathrm{concate}(f_{i2i},f_{q2q},f_{q2i},f_o)))$$
The objective function of the multi-interest network is the cross entropy loss function as follows:
$$L_{CTR} = -\frac{1}{N}\sum_{i=0}^N y_i {\mathrm{log}(\mathrm{pctr}_i)} + (1-y_i) {\mathrm{log}(1 - \mathrm{pctr}_i)}$$
Where $N$ is the total number of samples, $y_i$ is the ground truth label of the $i$th sample, and $\mathrm{pctr}_i$ is the CTR prediction from EGIN of the $i$th sample.

\subsection{End-to-End Joint Training}
 In most previous works, graph embedding learning takes a different form of input compared with the CTR prediction task. Also, they usually rely on graph engines to restore and sample graphs. These problems bring difficulties to end-to-end joint training of the two tasks. In previous sections, we introduce two key features of our proposed EGIN method: (i) light-weight graph sampling, (ii) unified input for graph and CTR prediction. Benefiting from the nature of EGIN, steam-style end-to-end training is effortless to implement in both offline and online scenarios. 

As shown in Figure~\ref{fig:framework}, both tasks receive the same raw input sampled from search impression logs, and the training objectives of different parts of EGIN are jointly considered for model optimization:
\begin{equation}
L=L_{CTR} + \alpha L_{i2i} + \beta L_{q2q}+ \gamma L_{q2i}
\label{eq:loss}
\end{equation}

Compared with two-stage methods that first train graph embedding and then exploit it in CTR prediction, the end-to-end framework comes with many advantages. It can avoid saving and loading graph embedding frequently, and therefore consistency between the two tasks is better guaranteed. With the gradient backpropagated from
both tasks, the graph embedding is jointly trained to fill the gap between graph learning and final CTR prediction.

\section{Experiments}
In this section, we conduct extensive experiments to demonstrate the effectiveness of our proposed methods. First, we evaluate the performance of different methods on public and industrial datasets. Second, we discuss our training detail and the corresponding influence. Third, we analyze the graph embedding result by visualization and statistics. Finally, we introduce our online A/B test result.

\subsection{Datasets}
Model comparisons are conducted on a commonly used public Taobao dataset \cite{Zhu_2018_taobao} and an industrial dataset collected from our e-commerce system. Table~\ref{tab:dataset} shows the statistics of two datasets.

\textbf{Taobao Dataset} \footnote{\url{https://tianchi.aliyun.com/dataset/649}} is a collection of user behaviors from Taobao's recommender system. The dataset contains four types of user behaviors, including click, purchase, adding to cart, and adding to wishlist. We take the click behaviors for each user and sort them according to time to construct the behavior sequence. Assuming there are $T$ behaviors of a user, we use the former $T-1$ clicked items as features to predict whether the user will click the $T$-th item. The behavior sequence is truncated at length 200.

\textbf{Industrial Dataset} is collected from our online search system, one of the world's largest e-commerce platforms. Samples are constructed from impression logs data of search results page. Each instance contains the user's historical click sequence and search query sequence with timestamps, as well as other user-side or item-side features, with "click" or "not" as the label. With the rich features contained, the collected industrial dataset allows us to build the item-query heterogeneous graph. The dataset is composed of training samples from the past 25 days and test samples from the following day, a classic setting for industrial modeling. The click and query sequences are truncated at length 100.

\begin{table}[t]
  \caption{Statistics of datasets used in paper}
  \label{tab:dataset}
  \begin{tabular}{lccc}
    \toprule
    Dataset & Sample & Items & Queries\\
    \midrule
    Taobao & 100 million & 4.16 million & None \\
    Industrial & 11.2 billion & 473 million & 228 million \\
  \bottomrule
\end{tabular}
\end{table}

\subsection{Compared Methods}
We compare EGIN with some mainstream CTR prediction methods, including three types: pooling-based, attention-based, and graph-based methods. The AUC is adopted as the performance metric, representing the ranking ability of the model. The same train and test data are used in all methods. 

EGIN adopts the following settings in experiments. The window size is set to 2 during the graph construction, and top-10 retrieval is used in the multi-interest network. The scale of the negative sampling queue is maintained at 1 million, and the negative sample number is set to 100. All embeddings in both graph and CTR prediction network share a dimension of 10.
\begin{itemize}
\item \textbf{DNN}. Sum-pooling is adopted on user behavior sequence to summarize the historical interest of users, which is concatenated with target item features, user features, and query features. Finally, they are fed into an MLP and get the CTR prediction.
\item \textbf{DNN-cross}. Compared with DNN, the cartesian product between historical and target items is adopted to represent the i2i relationships. Then this 2-order feature is fed into the sum-pooling layer to generate user interest representation. We called this "DNN-cross", and by default, the following model comparisons are based on this method.
\item \textbf{DIN} \cite{zhou2018deep}. DIN introduces an attention mechanism to assign different weights to historical items based on their relationships with the target item to learn the representation of user interests adaptively.
\item \textbf{BST} \cite{chen2019behavior}. BST utilizes Transformer to capture underlying sequential signals from user behavior sequences for a better recommendation.
\item \textbf{EGES} \cite{wang2018billion}. EGES constructs an item graph depending on user behaviors and then adopts DeepWalk to learn the representation of items. In order to alleviate the sparsity and cold start problems, side information is incorporated into the graph to enhance the embedding procedure. We consume the graph embedding produced by EGES in our CTR prediction network to evaluate its performance.
\end{itemize}

\subsection{Results on Public Dataset}

\begin{table}[t]
  \caption{Result on Taobao dataset}
  \label{tab:pub_result}
  \begin{tabular}{llcc}
    \toprule
    Method & AUC & RelaImpr\\
    \midrule
    DNN & $0.8709$ & $-0.16\%$\\
    DNN-cross & $0.8715$ & $0.00\%$\\
    DIN & $0.8833$ & $3.18\%$\\
    BST & $0.9040$ & $8.75\%$\\
    EGIN(ours) & $\mathbf{0.9184}$  & $\mathbf{12.62\%}$ \\
  \bottomrule
\end{tabular}
\end{table}

Considering the absence of query in the recommendation scenario of Taobao Dataset, we adjust our framework by preserving only nodes and edges of items in our query-item heterogeneous graph and eliminating features related to query in the multi-interest network.

Table~\ref{tab:pub_result} indicates that the proposed framework outperforms competitors on this commonly used CTR prediction benchmark. Our approach significantly achieves a better result than pooling-based and attention-based baselines. The result also reveals that our approach is a generalized CTR prediction framework that can be fit in recommendation scenarios.

\subsection{Results on Industrial Dataset}
\begin{table}[t]
  \caption{Result on industrial dataset}
  \label{tab:result}
  \begin{tabular}{llcc}
    \toprule
    Category & Method & AUC & RelaImpr\\
    \midrule
    \multirow{2}{*}{Pooling-based}  & DNN & $0.701$1 & $-0.30\%$\\
    & DNN-cross & 0.7017 & $0.00\%$\\
    \cline{1-4}
    \multirow{2}{*}{Attention-based} & DIN & $0.7022$ & $0.25\%$\\
    & BST & 0.7042 & $1.24\%$\\
    \cline{1-4}
    \multirow{2}{*}{Graph-based} & EGES & $0.7079$ & $3.07\%$\\
    & EGIN(ours) & $\mathbf{0.7108}$  & $\mathbf{4.51\%}$ \\
  \bottomrule
\end{tabular}
\end{table}

Table~\ref{tab:result} shows the results on our industrial dataset. The proposed approach achieves significant performance gain compared with DNN-based methods. Compared with the attention-based DIN and BST, we can see that incorporating graph learning can provide extra information to the CTR prediction model and improve model performance. In comparison with another graph-based method EGES, which utilizes co-occurrence and side information of commodity, our end-to-end framework shows a better result that proves the effectiveness of our query-item heterogeneous graph. Notice that our graph learning method brings low computation cost in addition to the main CTR prediction task.

\subsection{Ablation Study}
\begin{table}[t]
  \caption{Ablation study}
  \label{tab:ablation}
  \begin{tabular}{lcc}
    \toprule
    Method & AUC & Diff $\%$\\
    \midrule
    EGIN & 0.7108 &  \\
    EGIN w/o graph & 0.7032 & -0.76$\%$ \\
    EGIN w/o query & 0.7081 & -0.27$\%$ \\
    EGIN w/o pos\_emb & 0.7087 & -0.21$\%$ \\
  \bottomrule
\end{tabular}
\end{table}

The result of the ablation study on the industrial dataset is shown in Table~\ref{tab:ablation}. EGIN w/o graph removes the graph model and uses embedding produced by the CTR prediction network instead of jointly learned graph embedding. The significant performance decay indicates that the graph model provides plenty of extra information to the CTR network. EGIN w/o query means graph structure is limited to historical click sequence, causing a significant AUC drop. The result implies that query-item heterogeneous graph enhances CTR prediction performance by exploiting the information provided by queries. EGIN w/o pos\_emb eliminates the positional embedding in the multi-interest network, which brings the problem of order information loss during top-k retrieval. The AUC diff of $0.21\%$ conveys the necessity of this component in the network.

\subsection{Graph Input Data Management}
\subsubsection{Subsampling of Frequent Entites}
As our graph model shares the same input form with the CTR prediction network, it brings the problem of frequent items and queries. In our industrial dataset, the frequency of different items can vary from the range of ${10}^4$. In our general practice, subsampling of frequent entities helps improve performance in different search-related tasks. In order to verify the influence of this technique, we conduct a series of experiments on our industrial dataset by adopting different subsampling functions and subsampling targets (e.g., original behavior sequence or negative sample pool). In experiments, the subsampling method results in a CTR AUC of 0.7109 compared with 0.7108 for the unsampled version. The performance gap between different settings is limited to $10^{-4}$, which is negligible. 

\subsubsection{Combination of Categorically Restricted Data}

\begin{figure}[h]
  \centering
  \includegraphics[width=\linewidth]{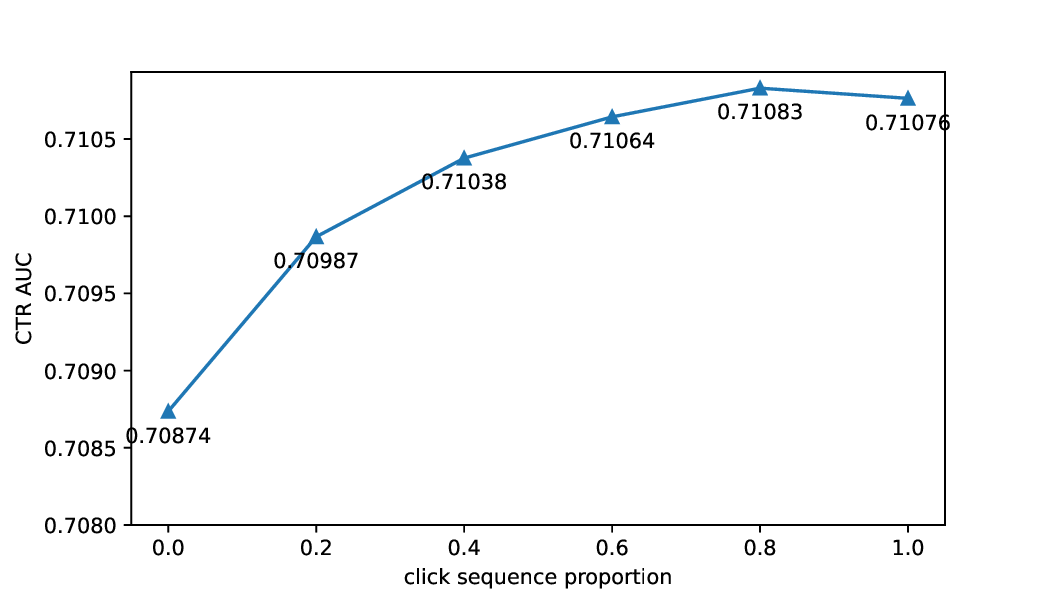}
  \caption{CTR prediction performance on industrial dataset of different mixtures of user click sequence and seeds sequence as input to graph learning. While the click sequence contains more inter-category transitions, the seeds sequence concentrates more on items corresponding to the search intention. $\mathbf{80\%}$ of click sequence and $\mathbf{20\%}$ of seeds sequence achieves best result in our experiment.}
  \Description{Similarity matrix.}
  \label{fig:seeds}
\end{figure}

In our search systems, a variant of user click sequence is available, and we call it seeds sequence. The seeds sequence is obtained by filtering out the items in user behavior sequence that belongs to different categories compared with the predicted intention of the current query. While click sequence offers more inter-category transfer information of user interest, seeds sequence concentrates more on intra-category similarity. In our query-item heterogeneous graph, the item2item relationship is expected to capture both types of information. Therefore, we adopt different mixture rates of click sequence and seeds sequence when building item2item relationship to test their performance. As shown in Figure~\ref{fig:seeds}, $80\%$ of click sequence and $20\%$ of seeds sequence are combined to achieve the best CTR AUC on the industrial datasets. Our approach adopts this proportion during other experiments for the best performance. 

\subsection{Analysis of EGIN Embedding}

\begin{figure}[h]
  \centering
  \includegraphics[width=\linewidth]{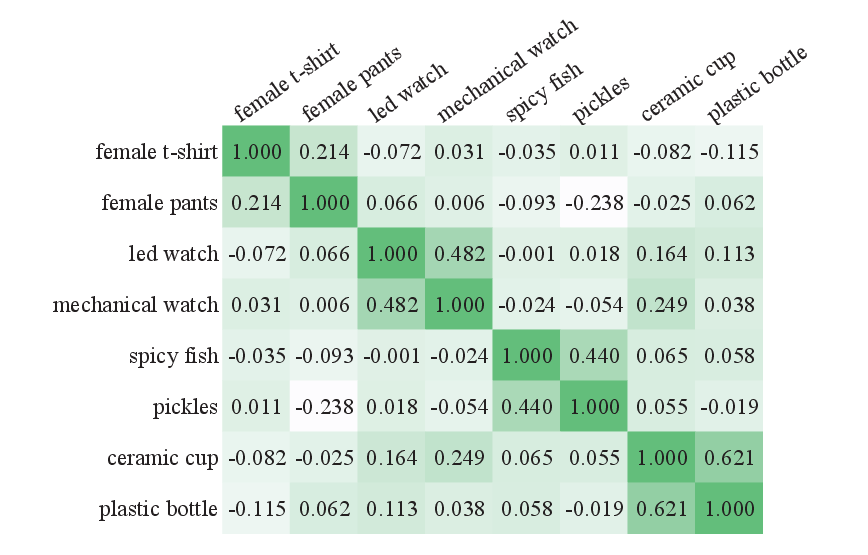}
  \caption{Graph embedding similarity matrix of several randomly selected item pairs from our industrial dataset.}
  \Description{Similarity matrix.}
  \label{fig:similarity}
\end{figure}

\begin{table}[t]
  \caption{Average similarity of EGIN and DNN embedding}
  \label{tab:similarity}
  \begin{tabular}{lccc}
    \toprule
    Average similarity & EGIN emb & EGES emb & DNN emb \\
    \midrule
    Intra-category & 0.6392 & 0.2973 & 0.0121 \\
    Inter-category & 0.0061 & 0.0121 & 0.0076 \\
  \bottomrule
\end{tabular}
\end{table}

To analyze the embedding learned by our EGIN model, we randomly choose four pairs of items from the industrial dataset to demonstrate their similarity. As shown in Figure~\ref{fig:similarity}, similarity within the pair is significantly higher. Belonging to different sub-categories, (female t-shirt, female pants) and (spicy fish, pickles) present similarities of 0.214 and 0.440, demonstrating that the embedding learned by EGIN is able to model various similarity information.

Besides, we randomly choose 1 million intra-category and inter-category item pairs from our industrial dataset and test their average cosine similarities of embeddings produced by different methods. Table~\ref{tab:similarity} shows us that the intra-category similarity of EGIN is much higher than the inter-category, whereas the difference from DNN embedding is relatively indistinctive. At the same time, our graph embedding appears to be more capable of capturing relationships between similar items than EGES and DNN.

\subsection{Online A/B Test}
We conduct online experiments in an A/B testing framework to further evaluate the performance of EGIN. The experimental goal is Click-Through-Rate (CTR) on the search result page of our e-commerce platform. The experiment lasts for 14 days and EGIN achieves a 2.76\% CTR gain compared to the product model, showing the great application value of the proposed approach.

\section{Conclusion}
In this paper, we propose a novel approach named EGIN for CTR prediction in e-commerce search scenarios. End-to-end joint training of graph embedding and CTR prediction is implemented in a light-weight framework without reorganization of graph data or dependency on graph engines. Our query-item heterogeneous graph well exploits impression log data of search systems to take query-item correlation and their sequential information into account to improve CTR prediction performance. Multi-interest network for CTR prediction is designed to comprehensively consume various information provided by item and query embeddings provided by the graph neural network. Offline experiments on public and industrial datasets demonstrate that the proposed EGIN outperforms attention-based and graph-based competitors. Online A/B test result reveals the great application value of the proposed approach.

Our work is an initial step towards a highly efficient heterogeneous graph learning method capturing various similarity information for CTR prediction in e-commerce search. The proposed framework has low computation cost and no dependency on the graph engine, which results in an effortless implementation in large-scale systems. At the same time, our method can be generalized to different search and recommendation systems and becomes a helpful plug-in for CTR prediction by providing high-quality embedding. In the future, we will integrate other types of nodes into the heterogeneous graph to complete the structure and improve the performance. We will also explore a better way to integrate graph embedding information in the CTR prediction network.

\bibliographystyle{ACM-Reference-Format}
\bibliography{main}

\end{document}